\documentclass[conference]{IEEEtran}
\IEEEoverridecommandlockouts

\usepackage{cite}
\usepackage{amsmath,amssymb,amsfonts}
\usepackage{algorithmic}
\usepackage{graphicx}
\usepackage{textcomp}
\usepackage{xcolor}
\def\BibTeX{{\rm B\kern-.05em{\sc i\kern-.025em b}\kern-.08em
    T\kern-.1667em\lower.7ex\hbox{E}\kern-.125emX}}
\makeatletter
\newcommand{\linebreakand}{%
  \end{@IEEEauthorhalign}
  \hfill\mbox{}\par
  \mbox{}\hfill\begin{@IEEEauthorhalign}
}
\makeatother
\begin{document}

\title{Optimizing Neural Speech Codec for Low-Bitrate Compression via Multi-Scale Encoding}

\author{
\IEEEauthorblockN{1\textsuperscript{st} Peiji Yang}
\IEEEauthorblockA{\textit{Interactive Entertainment Group} \\
\textit{Tencent }\\
Shenzhen, China \\
peijiyang@tencent.com}
\and
\IEEEauthorblockN{2\textsuperscript{nd} Fengping Wang}
\IEEEauthorblockA{\textit{School of Artificial Intelligence} \\
\textit{Beijing University of Posts and Telecommunications}\\
Beijing, China \\
wfp@bupt.edu.cn}
\and
\IEEEauthorblockN{3\textsuperscript{rd} Yicheng Zhong}
\IEEEauthorblockA{\textit{Interactive Entertainment Group} \\
\textit{Tencent}\\
Shenzhen, China \\
ajaxzhong@tencent.com}
\linebreakand 
\IEEEauthorblockN{4\textsuperscript{th} Huawei Wei}
\IEEEauthorblockA{\textit{Interactive Entertainment Group} \\
\textit{Tencent}\\
Shenzhen, China \\
huaweiwei@tencent.com}
\and
\IEEEauthorblockN{5\textsuperscript{th} Zhisheng Wang}
\IEEEauthorblockA{\textit{Interactive Entertainment Group} \\
\textit{Tencent}\\
Shenzhen, China \\
plorywang@tencent.com}
}

\maketitle
\begin{abstract}
Neural speech codecs have demonstrated their ability to compress high-quality speech and audio by converting them into discrete token representations. 
Most existing methods utilize Residual Vector Quantization (RVQ) to encode speech into multiple layers of discrete codes with uniform time scales. However, this strategy overlooks the differences in information density across various speech features, leading to redundant encoding of sparse information, which limits the performance of these methods at low bitrate.
This paper proposes MsCodec, a novel multi-scale neural speech codec that encodes speech into multiple layers of discrete codes, each corresponding to a different time scale. This encourages the model to decouple speech features according to their diverse information densities, consequently enhancing the performance of speech compression. Furthermore, we incorporate mutual information loss to augment the diversity among speech codes across different layers.
Experimental results indicate that our proposed method significantly improves codec performance at low bitrate.

\end{abstract}
\begin{figure*}[htbp]
	\centering
        \setlength{\abovecaptionskip}{0.3cm}
	\includegraphics[width=\linewidth]{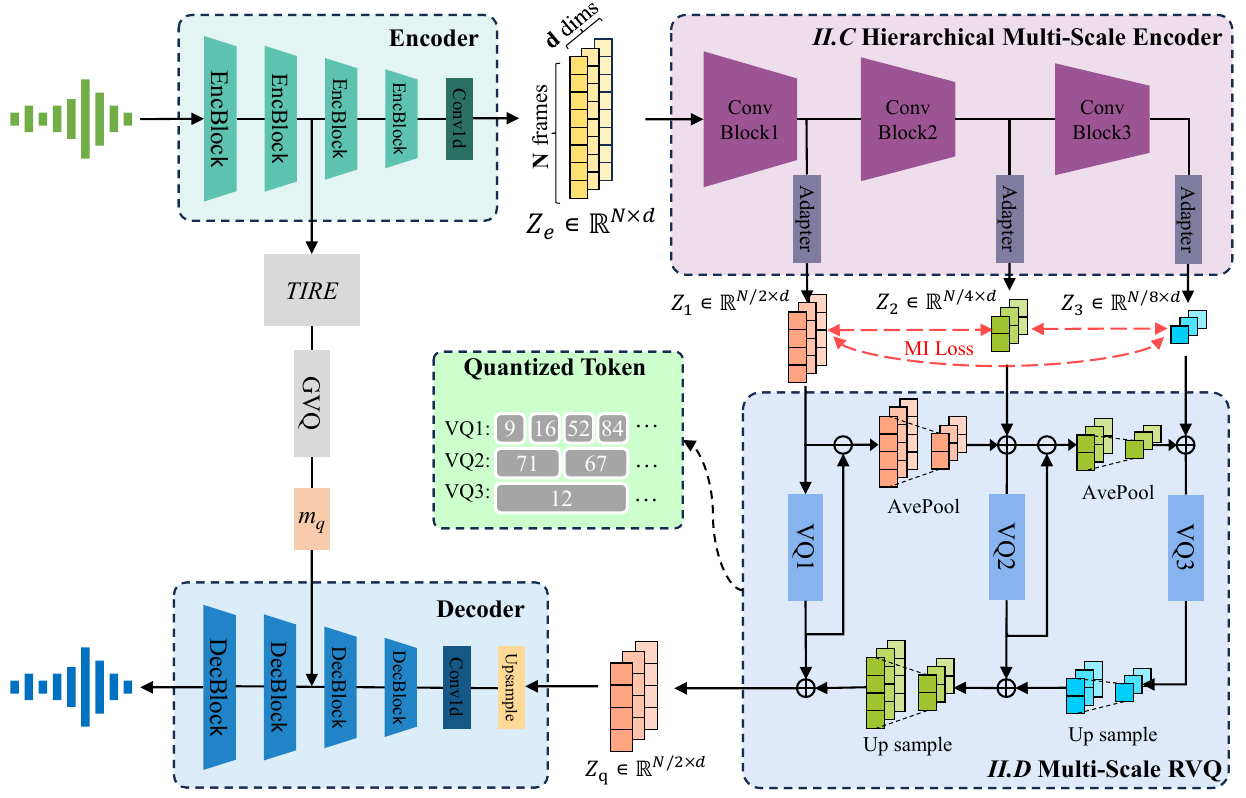}
	\caption{The architecture of MsCodec. }
	\label{fig:overall}
        \vspace{-0.5cm}
\end{figure*}
\begin{IEEEkeywords}
neural speech codec, low bitrate, multi-scale
\end{IEEEkeywords}
\section{Introduction}

In recent years, neural speech codecs~\cite{zeghidour2021soundstream, defossez2022high, zhang2023speechtokenizer,du2024funcodec, yang2023hifi} have garnered significant attention due to their ability to effectively compress high-quality speech and audio by converting them into discrete tokens.
Efficient audio compression is crucial for various applications, including transmission in bandwidth-limited environments~\cite{kolundvzija2024low} and low-latency communication systems~\cite{kondoz2005digital}. Moreover, neural speech codecs are essential to developing speech synthesis based on the Language Models~\cite{wang2023neural, kharitonov2023speak}, enabling high-quality and natural-sounding speech generation. 
These applications require codecs to deliver high-quality speech at low bitrate, but this remains an unresolved issue.

The RVQ-based models are highly competitive in discrete codecs, typically comprising an encoder, an RVQ module, and a decoder.
The encoder transforms the audio into a sequence of speech embeddings. The RVQ module then replaces each embedding with a sum of vectors from a finite set of codebooks. Following this, the decoder reconstructs the audio from the quantized embeddings. Within the RVQ process, a series of quantizers are employed, with each quantizer responsible for quantizing the residual from the previous one. 
While this method effectively reduces the gap between quantized embeddings and original speech embeddings when sufficient codebooks are available, solely quantizing residuals introduces redundancy. 
Consequently, its performance falls short of expectations when the number of codes is limited.
 
Recent studies have focused on decoupling speech attributes to enhance codec performance.
FACodec~\cite{ju2024naturalspeech} employs additional supervisory signals related to speech attributes to facilitate the decoupling of speech information. 
However, it lacks optimization for low bitrate scenarios and relies heavily on external models, which complicates the training process. 
TiCodec~\cite{ren2024fewer} and SingleCodec~\cite{li2024single} incorporate time-invariant information, decoupling speech features that remain constant over time. This implicit decoupling method reduces the encoding burden.
While these methods, through explicit or implicit decoupling, have improved codec performance, none account for the varying information densities of speech features, leading to the persistence of redundancy issues.

To address these issues, we propose MsCodec which can capture speech features with varying information densities. 
Specifically, information density is related to the variation of speech features at different time scales.
Building on this, we developed a Multi-Scale Hierarchical Encoder that encodes speech into embeddings at different time scales, enabling the model to autonomously capture various speech features. Additionally, to minimize redundancy across the hierarchical encoder layers, we apply mutual information~\cite{cheng2020club} to reduce the correlation between different hierarchical embeddings.
To quantize these multi-scale embeddings, we further propose the Multi-Scale RVQ module, 
which can quantize embeddings at different scales while following the RVQ strategy to reduce quantization errors.
Finally, the quantized embeddings are reconstructed into audio by the Decoder.
Comprehensive experiments validate the effectiveness of each proposed module and demonstrate that our method outperforms the baseline at low bitrate. Some audio samples are availabe\footnote{https://tencentgamemate.github.io/MsCodec-Demo/}.
\section{Methodology}
\subsection{Overview}
The architecture of MsCodec is shown in Figure \ref{fig:overall}. We consider a single-channel speech as ${x} \in \mathbb{R}^T$. An Encoder transforms $x$ into a sequence of embeddings $ Z_e \in \mathbb{R}^{N \times d} $, where \( N \) is the number of embedding frames, which is determined by the Encoder's downsampling factor, and \( d \) denotes the hidden size of the embeddings. Following TiCodec, we incorporate a time-invariant representation extraction module, which processes the intermediate representation of the second convolutional module of the Encoder to produce a time-invariant representation \( m \in \mathbb{R}^d \), subsequently quantized using a group vector quantizer (GVQ)~\cite{yang2023hifi}, resulting in $m_q \in \mathbb{R}^d$. To capture speech representation at different time scales, a Hierarchical Multi-Scale Encoder is applied to further encodes $ Z_e $, resulting in representations at various scales: $ Z_{1} \in \mathbb{R}^{N/2 \times d} $, $ Z_{2} \in \mathbb{R}^{N/4 \times d} $, and $ Z_{3} \in \mathbb{R}^{N/8 \times d} $. These multi-scale representations are then fed into a Multi-Scale RVQ module for quantization, producing a compressed representation $ Z_q \in \mathbb{R}^{N/2 \times d} $. Finally, the Decoder reconstructs the speech $ \hat{x} $ from the compressed representations $Z_q$ and $m_q$. 

\subsection{Encoder \& Decoder Architecture}
The architecture of the Encoder and Decoder is inspired by Encodec.
The Encoder consists of a 1D convolutional layer followed by four convolutional modules and a final 1D convolutional layer. Each convolutional module consists of three residual units and one downsampling layer. 
Together, these four modules result in a total downsampling factor of \( D \). For instance, with strides of \( (2, 4, 5, 8) \), the total downsampling is calculated as $D = 2 \times 4 \times 5 \times 8 = 320$, which determines the number of frames \( N \) for \( Z_e \).
The Decoder begins with an additional convolutional module to upsample the input $Z_q$, after which the remaining structure mirrors the encoder, utilizing transpose convolutions for upsampling.

\subsection{Hierachical Multi-Scale Encoder}
The Hierarchical Multi-Scale Encoder is designed to capture and encode features from speech signals at varying time scales.
The hierarchical encoder consists of three convolutional modules, each comprising three residual units and a downsampling layer. With strides set to (2,2,2), the frame length of the embeddings is halved after each convolutional module, progressively encoding $Z_e$ at coarser time scales. The output of each convolutional module not only serves as the input for the next convolutional module but is also processed through a corresponding adapter to obtain new embeddings at different scales. Each adapter is composed of two Conformer~\cite{gulati2020conformer} layers. 

The Hierarchical Multi-Scale Encoder generates the representations $Z_{1},Z_{2},Z_{3}$, each capturing different aspects of the speech signal. Specifically, $Z_{1}$ captures rapid fluctuations and fine-grained details at the finest scale, $Z_{2}$ balances between fine and coarse features at an intermediate scale, and $Z_{3}$ focuses on long-term dependencies at the coarsest scale.

\subsection{Multi-Scale RVQ}
To efficiently compress speech embeddings across multiple scales and minimize compression-induced errors, we propose a Multi-Scale RVQ module.
The process begins by quantizing the input \( Z_{1} \) at the finest time scale using a vector quantizer $VQ_1$, producing the quantized embedding \( \hat{Z_{1}} \). The residual between \( \hat{Z_{1}} \) and \( Z_{1} \) is aligned with \( Z_{2} \) through mean pooling and then added to $Z_{2}$. The updated $Z_{2}$ is fed into the second vector quantizer $VQ_2$ for quantization, producing $\hat{Z_{2}}$. The same process is applied to $Z_{3}$, producing $\hat{Z_{3}}$. 
Finally, $\hat{Z_2}$ and $\hat{Z_3}$ are aligned with $\hat{Z_1}$ in the time dimension by frame-wise duplication and then summed to obtain the compressed representation $Z_q$. Therefore, the Multi-Scale module can not only quantize embeddings at different scales but also reduce compressed loss through a specialized residual technique.


\subsection{Mutual Information Loss}
The embeddings of different time scales are extracted from the same speech, which can lead to a certain degree of information redundancy. Mutual information is employed to further decouple these representations to address this issue. To estimate mutual information, we utilize the Variational Contrastive Log-Upper Bound (vCLUB) method~\cite{cheng2020club,wang2021vqmivc}, which offers an upper bound for it. Given the variables u and v, the vCLUB formulation for mutual information is given by:
\begin{equation}\hspace{-2mm}
I(u, v) \!=\! \mathbb{E}_{P(u,v)} \left[ \log Q_{\theta}(u|v) \right]\!-\!\mathbb{E}_{P(u)} \mathbb{E}_{P(v)} \left[ \log Q_{\theta}(u|v) \right] 
\end{equation}
$Q_{\theta}(u|v)$ is the variational approximation which makes vCLUB a reliable MI estimator. The variational approximation network is computed as follows:
\begin{equation}
    \mathcal{L}_{u,v} = \log Q_{\theta_{u,v}}(u | v)
\end{equation}
In the training process, we sample \( K \) examples to form a batch. To facilitate direct comparison across different time scales, we extend \( Z_{2} \) and \( Z_{3} \) to match the frames of \( Z_{1} \). Specifically, \( Z_{2} \) is repeated frame-wise by a factor of two, and \( Z_{3} \) by a factor of four. With these aligned representations, the estimation for vCLUB between different time-scale representations is given by:
\begin{equation}
\begin{aligned}\hspace{-1mm}
\hat{I}(Z_{1},Z_{2}) \!&=\! \frac{2}{K^2 N} \sum_{k=1}^{K} \sum_{l=1}^{K} \sum_{t=1}^{N/2} \\
&\quad \left[ \log Q_{\theta_{1,2}} (Z_{1}^{k,t} | Z_{2}^{k,t}) \!-\! \log Q_{\theta_{1,2}} (Z_{1}^{l,t} | Z_{2}^{k,t}) \right]
\end{aligned}
\end{equation}
Therefore, we can decrease the correlation among different time-scale representations by the MI Loss:
\begin{equation}
    \mathcal{L}_{mi} = \hat{I}(Z_{1},Z_{2}) + \hat{I}(Z_{1},Z_{3}) + \hat{I}(Z_{2},Z_{3}) 
\end{equation}

Finally, our model is trained to optimize the following composite loss function:
\begin{equation}
    \mathcal{L} = \lambda_t \mathcal{L}_t + \lambda_f \mathcal{L}_f + \lambda_g \mathcal{L}_g + \lambda_{\text{feat}} \mathcal{L}_{\text{feat}} + \lambda_{w} \mathcal{L}_{w} + \lambda_{mi} \mathcal{L}_{mi}
\end{equation}
Where $\mathcal{L}_t$, $\mathcal{L}_f$, $ \mathcal{L}_g$, $\mathcal{L}_{\text{feat}}$ and $\mathcal{L}_{w}$ follow the losses used in Encodec. The hyperparameters \( \lambda_t, \lambda_f, \lambda_g, \lambda_{\text{feat}}, \lambda_{w}, \lambda_{mi} \) are used to balance each term in the total loss function. 
\begin{table*}[htp]
\centering
\small
\renewcommand{\arraystretch}{1.3}
\caption{Objective metrics scores of various codecs}
\label{tab: objective metrics}
\begin{tabular}{c c c c c c c c}
\hline
\textbf{Model}  & \textbf{Nq} & \textbf{Bandwidth(bps)↓}   & \textbf{STOI↑} & \textbf{PESQ↑} & \textbf{MCD↓} & \textbf{UTMOS↑} & \textbf{SPK↑} \\ \hline
Encodec-1VQ     & 1 & 750   & 0.721	&	1.357	&	4.570	&	2.438	&	0.779 \\
HiFiCodec-1VQ  & 1 & 750   & 0.726	&	1.354	&	4.508	&	2.521	&	0.749      \\
TiCodec-1VQ     & 1 & 750   & 0.737	&	1.371	&	4.405	&	2.496	&	0.760                      \\      
MsCodec-S       & 3 & \textbf{700}  & \textbf{0.753}	&	\textbf{1.485}	&	\textbf{4.291}	&	\textbf{2.570}	&	\textbf{0.793} \\ \hline
Encodec-2VQ     & 2 & 1500  &   0.804	&	1.692	&	3.622	&	2.734	&	0.819             \\
HiFiCodec-1VQ  & 2 & 1500  &   0.812	&	1.723	&	3.635	&	2.567	&	0.808      \\
TiCodec-2VQ     & 2 & 1500  &   0.817	&	1.730	&	3.513	&	2.617	&	0.811                 \\   
MsCodec-M       & 3 & \textbf{1400} &   \textbf{0.832}	&	\textbf{1.814}	&	\textbf{3.411}	&	\textbf{2.801}	&	\textbf{0.847}    \\ \hline
Encodec-4VQ     & 4 & 3000  &  0.847	&	2.029	&	3.175	&	2.832	&	0.833            \\
HiFiCodec-1VQ  & 4 & 3000  &   0.859	&	2.072	&	3.066	&	2.711	&	0.858      \\
TiCodec-4VQ     & 4 & 3000  &   0.862	&	2.020	&	2.997	&	2.706	&	0.845              \\ 
MsCodec-L       & 3 & \textbf{2800} &  \textbf{0.865}	&	\textbf{2.125}	&	\textbf{2.930}	&	\textbf{2.986}	&	\textbf{0.858} \\ \hline
\end{tabular}
\vspace{-0.3cm}
\end{table*}

\section{Expertiments}
\subsection{Experiments Setup} 
\noindent\textbf{Dataset.}
We trained speech codecs using five open-source datasets: LibriTTS~\cite{zen2019libritts}, Hi-Fi TTS~\cite{bakhturina2021hi}, VCTK~\cite{liu2019cross}, AISHELL-1~\cite{bu2017aishell}, and AISHELL-3~\cite{shi2020aishell}. 
The training set, totaling 1,113 hours of audio, was created by combining the training subsets from each dataset. For evaluation, we randomly sampled 200 utterances from the test subsets of these datasets. All audio data were resampled to 24 kHz.

\noindent \textbf{Baselines.}
We used Encodec, TiCodec, and HiFiCodec as baselines for comparison with our proposed MsCodec. Notably, TiCodec, HiFiCodec, and our MsCodec are all designed for lower bitrate compression.

\noindent \textbf{Training Details.}
We trained all codec models on 4 V100 GPUs for 300k iterations with a learning rate of 4e-4 and a batch size of 8 per GPU. For MsCodec, the codebook size is set to 1024, with a hidden size of 512. The intermediate hidden state sizes in the convolutional blocks of the Encoder are 128, 256, 512, and 1024. In the Hierarchical Multi-Scale Encoder, all convolutional blocks have a hidden size of 512, and the adapter also has a hidden size of 512. We set $\lambda_t=0.1, \lambda_f=1, \lambda_g=3, \lambda_{\text{feat}}=5, \lambda_{w}=1, \lambda_{mi}=0.01$. For Encodec and HiFiCodec, we use the implementation from AcademiCodec\footnote{https://github.com/yangdongchao/AcademiCodec}, while for TiCodec\footnote{https://github.com/y-ren16/TiCodec}, we utilize the official code.

\noindent \textbf{Evaluation Metrics.}
We evaluate the objective quality of the reconstructed speech using several metrics, including STOI~\cite{taal2010short}, PESQ~\cite{recommendation2001perceptual}, Mel Cepstral Distortion (MCD)~\cite{kubichek1993mel}, UTMOS~\cite{saeki2022utmos}, and speaker cosine similarity (SIM)~\cite{wang2024advancing}.

\subsection{Speech Reconstruction Performance}
To evaluate the speech reconstruction quality of MsCodec, we compared it with the baseline models Encodec, HiFiCodec, and TiCodec. These baseline models use 1, 2, and 4 quantizers. The compressed bitrate of MsCodec was adjusted by modifying the convolutional block strides in the Encoder to align the bitrates with these comparison models. Specifically, we designed three variants of MsCodec: MsCodec-S, MsCodec-M, and MsCodec-L. The stride configurations for these variants are MsCodec-S (5, 5, 4, 3), MsCodec-M (5, 5, 3, 2), and MsCodec-L (5, 5, 3, 1), corresponding to bitrates of 700 bps, 1400 bps, and 2800 bps, respectively.

Table \ref{tab: objective metrics} shows that MsCodec outperforms the baseline models across all objective metrics. Notably, at lower bitrate, MsCodec demonstrates a significant advantage.
This result aligns with our intuition. RVQ focuses solely on quantizing the residuals from the previous quantizer and encodes all speech information into embeddings at the same time scale, which inevitably leads to redundant encoding of sparse information. Consequently, when the number of codes is limited, it becomes increasingly difficult to capture all the information in the original speech.
Our proposed Multi-Scale strategy encodes speech into embeddings at different time scales based on the varying information densities of speech features, effectively reducing redundant encoding. This strategy significantly improves codec performance, especially when the number of codes is limited.

\subsection{Ablation Study}
For the ablation study, we selected MsCodec-M as the baseline model to investigate the contributions of the multi-scale strategy and mutual information.

\noindent \textbf{w/o MI Loss.} To assess the role of mutual information, we compared the performance of MsCodec-M with and without the MI Loss term. 

\noindent \textbf{w/o Multi-Scale. } To ensure that the performance improvement is not simply due to the increased number of parameters in the hierarchical encoder, we retained the hierarchical structure but set the strides of the hierarchical encoder to (1, 1, 1). This forces the model to generate a fixed-scale embedding, removing the multi-scale capability. Additionally, we adjusted the strides of the encoder to (8, 5, 4, 3), resulting in a bitrate of 1500 bps. We refer to this modified model as fix-scale codec medium (FsCodec-M). 

\begin{table}[htp]
    \centering
    \setlength{\abovecaptionskip}{0.1cm}
    \caption{Ablation Study}
    \renewcommand{\arraystretch}{1.2}
    \begin{tabular}{c c c c c}  
    \hline
         \textbf{Model} &  \textbf{Bandwidth(bps)} & \textbf{STOI↑} & \textbf{PESQ↑} & \textbf{MCD↓}  \\ \hline
         MsCodec-M &      1400  & \textbf{0.832}	&	\textbf{1.814}	&	\textbf{3.411}\\
         w/o MI Loss   &   1400  & 0.822	&	1.750	&	3.486 \\
         w/o Multi-Scale   &  1500  & 0.791 & 1.642 & 3.938 \\  \hline
    \end{tabular}
    \label{tab:ablation}
\end{table}

The results of these ablation experiments, as shown in Table \ref{tab:ablation}, demonstrate that both the multi-scale strategy and mutual information loss play significant roles in improving the performance of MsCodec.

\subsection{Analysis of Multi-Scale Strategy}
\begin{figure}[htbp]
	\centering
        \setlength{\abovecaptionskip}{0.cm}
	\includegraphics[width=\linewidth]{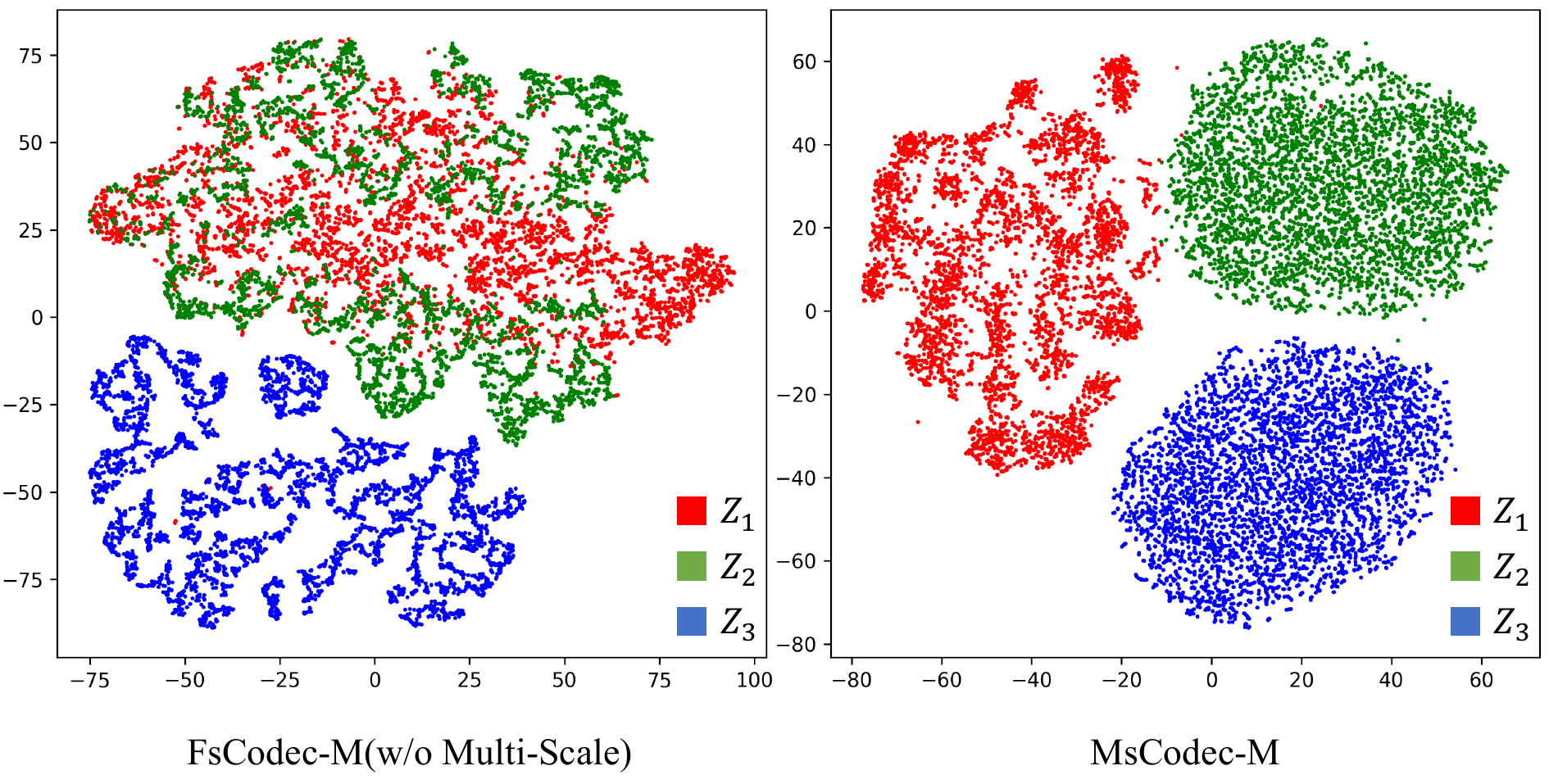}
	\caption{t-SNE of $Z_1$, $Z_2$, and $Z_3$}
	\label{fig:tsne}
        \vspace{-0.3cm}
\end{figure}

To validate the effectiveness of the Multi-Scale strategy in promoting the model’s ability to decouple speech features with varying information densities, we extracted the embeddings \( Z_1 \), \( Z_2 \), and \( Z_3 \) from both MsCodec-M and FsCodec-M. These embeddings were projected onto a 2D space using t-SNE for visualization~\cite{van2008visualizing}. As shown in Figure \ref{fig:tsne}, the distributions of \( Z_1 \), \( Z_2 \), and \( Z_3 \) in MsCodec-M are well-separated, indicating that the Multi-Scale strategy significantly aids in the self-decoupling of speech representations. In contrast, the distributions of \( Z_1 \) and \( Z_2 \) in FsCodec-M are entangled, likely because \( Z_1 \), \( Z_2 \), and \( Z_3 \) are encoded at the same time scale, resulting in representation collapse.

\section{conclusions}
In this paper, we proposed MsCodec, a multi-scale neural speech codec designed to enhance codec performance at low bitrate.
MsCodec effectively captures speech features across various time scales, reducing the issue of redundant information encoding present in previous methods, resulting in improved performance, especially at lower bitrate.
Through comprehensive experiments and analysis, we have validated the effectiveness of each proposed module and demonstrated that the multi-scale strategy significantly enhances the model's capacity to autonomously decouple speech features.

\newpage
\bibliographystyle{IEEEtran}
\bibliography{mybib}
\end{document}